\documentclass[twocolumns]{aa}  

\usepackage{graphicx}

\usepackage{float}
\usepackage{subfig}
\usepackage{txfonts}
\usepackage{url}
\usepackage[switch]{lineno}
\usepackage{datetime}
\usepackage{wrapfig}
\usepackage[usenames,table,dvipsnames]{xcolor}
\usepackage{setspace}
\usepackage[normalem]{ulem}
\usepackage{soul}
\usepackage{hyperref}
\usepackage{xspace}

\newcount\Comments  
\newcommand{\kibitz}[2]{\ifnum\Comments=1\textcolor{#1}{#2}\fi}

\definecolor{blue}{RGB}{66, 153, 233}
\definecolor{red}{RGB}{255, 0, 0}
\definecolor{purple}{RGB}{255, 0, 255}

\newcommand{\rxj}{RX J1713.7-3946\xspace}

\begin{document}

   \title{Mapping thermal emission in the synchrotron-dominated SNRs G330.2+1.0, 3C58, and \rxj}

    \author{A. Picquenot
          \inst{1,2,3}
          \and
          B.~J. Williams
          \inst{2}
          \and
          F. Acero
          \inst{4,5}
          \and
          K. Mori
          \inst{6,7}
          }

\institute{
              Department of Astronomy, University of Maryland, College Park, MD 20742 
              \and
              X-ray Astrophysics Laboratory NASA/GSFC, Greenbelt, MD 20771 
              \and
              Center for Research and Exploration in Space Science and Technology, NASA/GSFC, Greenbelt, MD 20771, USA
              \and
              Universit\'e Paris-Saclay, Universit\'e Paris Cit\'e, CEA, CNRS, AIM, 91191 Gif-sur-Yvette, France
              \and
              FSLAC IRL 2009, CNRS/IAC, La Laguna, Tenerife, Spain
              \and
              Department of Applied Physics and Electronic Engineering, University of Miyazaki, 1-1, Gakuen Kibanadai-nishi, Miyazaki 889-2192, Japan
              \and
              Institute of Space and Astronautical Science (ISAS), Japan Aerospace Exploration Agency (JAXA), Japan
}

   \date{\today}

 
  \abstract
   {}
   {Since the discovery of synchrotron X-ray emission from the shell of the supernova remnant (SNR) SN 1006, multiple observations from {\it Chandra} and {\it XMM-Newton} have shown that many young SNRs produce synchrotron emission in X-rays. Among those, a few peculiar SNRs have their X-ray emission largely dominated by synchrotron radiation, showing no or only faint traces of thermal emission. In this paper, we report our mapping of the thermal emission in three emblematic synchrotron-dominated SNRs: G330.2+1.0, 3C58, and \rxj.}
   {We used a blind source separation method able to retrieve faint components from X-ray data in the form of {\it Chandra} and {\it XMM-Newton} observations. The thermal candidates disentangled by the algorithm were then used to select regions of extraction. We then analyzed the extracted spectra to assess their physical nature.}
   {We conclude that the components retrieved by the algorithm indeed represent the spatial distribution of the thermal emission in G330.2+1.0 and 3C58, and a likely thermal candidate in \rxj. Our findings confirm and expand on past studies. }
   {}

   \keywords{ }

   \maketitle
%

\section{Introduction}
\label{sect:intro}

Since the discovery of synchrotron X-ray emission from the shell of SN 1006 \citep{Koyama1995}, multiple observations from {\it Chandra} and {\it XMM-Newton} have shown that many young supernova remnants (SNRs) produce synchrotron emission in X-rays. Among those, a few peculiar SNRs have their X-ray emission largely dominated by synchrotron radiation, showing no or only faint traces of thermal emission.

Of these synchrotron-dominated SNRs, the remnant of SN 1054, dominated by the pulsar wind nebula (PWN) know as  the Crab, is the most notorious example: despite numerous attempts \citep[e.g.,][]{Seward_2006,hitomicollab}, no evidence of thermal emission has ever been found. G330.2+1.0, the SNR surrounding 3C58, and \rxj are emblematic synchrotron-dominated SNRs in which traces of thermal emission have been detected in previous studies.

G330.2+1.0 is a synchrotron-dominated SNR showing a clear shell in X-rays, and harbors a central compact object (CCO) \citep{Park_2006}, presumably a neutron star. Its age has been estimated to be less than about 1000 yr by \cite{Borkowski_2018}.

The PWN 3C58 harbors a 66 msec pulsar in its center, and presents axisymmetric lobes. It has been linked to the SN explosion observed in A.D. 1181 by Chinese and Japanese astronomers \citep{1971QJRAS..12...10S}, but it has also been argued that some of its characteristics were indicative of a much older age \citep[see e.g.,][]{1993A&A...274..427R}. However, a new distance estimate of $2$ kpc instead of the traditionally accepted $3.2$ kpc would be consistent with an age of $\sim 840$ yr \citep{Kothes13}. 

\rxj is a shell-type SNR, and the brightest
X-ray synchrotron and TeV gamma-ray SNR in our Galaxy. \citet{1997A&A...318L..59W} proposed that \rxj could be the remnant of the SN that exploded in A.D. 393.

In \cite{Park_2009} and  \cite{Williams_2018}, a small region to the east of G330.2+1.0 was found to be a source of thermal emission. \cite{Bocchino01}, \cite{Slane04}, and \cite{Gotthelf_2007} detected evidence of thermal emission in 3C58, the latter producing a synchrotron-subtracted image of the remnant using stacked {\it XMM-Newton} observations. Finally, thermal emission was detected in a central region of \rxj by \cite{Katsuda15} after numerous unsuccessful or debated attempts \citep[e.g.,][]{Koyama97,Slane_1999,Pannuti_2003}. However, these past studies mainly focused on the spectral analysis of small regions within the remnants and, with the exception of the synchrotron-subtracted map of  \cite{Gotthelf_2007}, there was no real attempt to probe the global spatial distribution of thermal emission in these SNRs. 

In this paper, we propose a different approach to tackling the problem: we attempt to map the thermal emission in G330.2+1.0, 3C58, and \rxj with a new method. In Section \ref{sect:method}, we present our image-extraction method, while the following sections present our results for G330.2+1.0, 3C58, and \rxj, respectively.

\begin{table*}
    \centering

    \renewcommand{\arraystretch}{1.4}
    \begin{tabular}{c c c c c c c}

     & & &  & Exposure (ks) &  &\\ \cline{4-7}
   Object & ObsID & Date & MOS 1 & MOS 2 & pn & ACIS-S  \\ 
    \hline 
 G330 & 0742050101 & 2015 Mar 8 \ & 112.4 & 109.3 & 111.2 & -
 \\ 
\hline
3C58 & 728 & 2000 Sep 4 \ & - & - & - & 50.0
 \\ 
  & 3832 & 2003 Apr 26 \ & - & - & - & 135.8
 \\ 
   & 4382 & 2003 Apr 23 \ & - & - & - & 167.4
 \\ 
   & 4383 & 2003 Apr 22 \ & - & - & - & 38.7
 \\ 
 \cline{3-7}
  & & Total & - & - & - & 391.9\\
\hline 

RX J1713 & 0093670501 & 2001-03-02 & 11.6 & 11.6 & 6.5 & -\\
& 0203470501 & 2004 Mar 25 & 13.6 & 13.6 & 9.4 & -\\
& 0740830201 & 2014 Mar 2 & 89.5 & 89.5 & - & -\\
& 0743030101 & 2015 Mar 10 & 67.0 & 67.0 & 38.3 & -\\ 
& 0804300901 & 2017 Aug 29 & 19.5 & 19.5 & 14.2 & -\\ 
& 0804300801 & 2017 Aug 30 & 43.6 & 43.6 & 33.0 & -\\ 
& 0804301001 & 2018 Mar 23 & 50.6 & 52.3 & 32.2 & -\\ 
& 0804300301 & 2018 Mar 29 & 55.7 & 55.7 & 34.1 & -\\ 
& 0804300401 & 2018 Mar 31 & 51.5 & 51.5 & 28.0 & -\\ 
& 0804300501 & 2018 Mar 25 & 97.5 & 97.6 & 55.1 & -\\ 
& 0804300601 & 2018 Mar 19 & 49.8 & 49.8 & 38.7 & -\\ 
& 0804300701 & 2018 Mar 21 & 70.7 & 70.7 & 41.4 & -\\ 
& 0804300101 & 2018 Aug 26 & 69.7 & 69.7 & 53.7 & -\\ 
 \cline{3-7}
  & & Total & 690.3 & 692.1 & 384.6 & -\\
\hline 
\end{tabular}
\caption{\label{tab:data-description}Data from {\it XMM-Newton} and {\it Chandra} used in our study. EPIC exposure times are after filtering.}
\label{sect:obs}
\end{table*}

\section{Method}
\label{sect:method}

To retrieve accurate maps of the thermal emission in the three synchrotron-dominated SNRs we propose to study, we used a blind source separation (BSS) method based on the general morphological components analysis \citep[GMCA; see][]{bobin15} first introduced for X-ray observations by \cite{picquenot:hal-02160434}. This algorithm can disentangle spectrally and spatially mixed components from an X-ray data cube of the form $(x,y,E)$. In particular, the algorithm was shown to be able to extract extremely faint components from X-ray data cubes. The outputs take the form of an image associated with a mean spectrum for each component. As the algorithm primarily focuses on spatial morphological features without the help of any prior physical information, the extracted spectra can sometimes appear poorly reconstructed, particularly for faint components. These spectra can nonetheless provide hints regarding the nature of the extracted component, and the images can be used to extract a spectrum from the original data in the region of interest. The GMCA algorithm can therefore be used as an effective complimentary tool for the search for faint emission within spatially resolute sources. An updated version, the pGMCA \citep[see][]{9215040}, has been developed to take into account the Poissonian nature of X-ray data. It was first used on Cas~A SNR data from {\it Chandra} and proved to be perfectly suited for producing clear, detailed, and unpolluted images of both the ejecta \citep{Picquenot_2021} and the synchrotron at different energies \citep{Picquenot_2023}. 

In the present study, we used pGMCA on stacked {\it Chandra} observations of 3C58 and, for the first time, on an {\it XMM-Newton} observation of G330.2+1.0. As the algorithm does not handle mosaic data with highly uneven exposure maps, we combined {\it XMM-Newton} observations of the large \rxj remnants, corrected the exposure, and removed the background and solar flares to generate a flux data cube. We also removed the CCOs from all the remnants with an inpainting method using wavelet transforms in order to alleviate the signal intensity contrast.

\section{G330.2+1.0}
\label{sect:G330}

\begin{figure*}[ht!]
\centering
\subfloat{\includegraphics[width = 18cm]{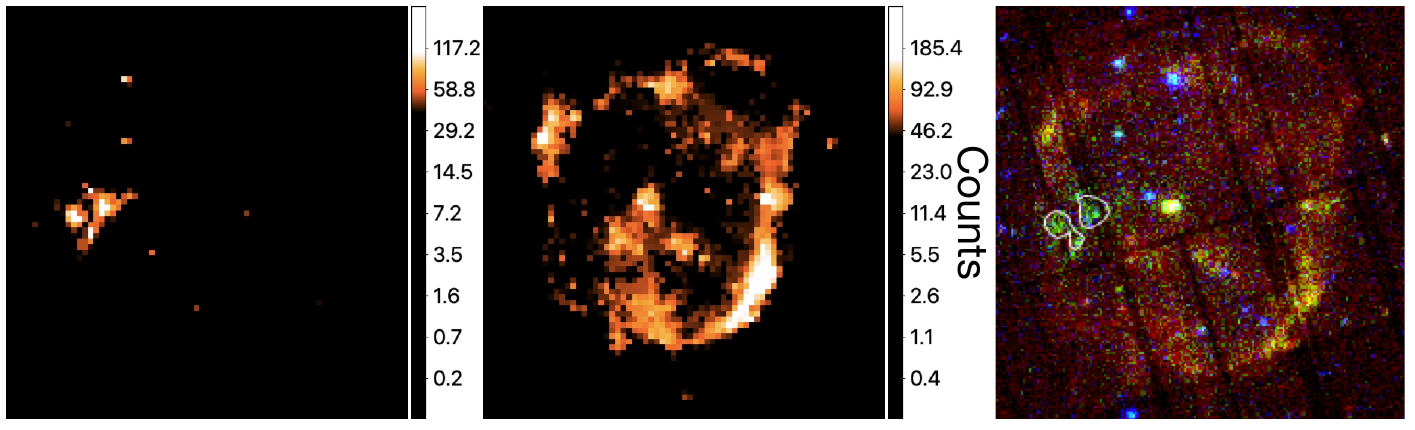}}
\caption{Thermal emission distribution in G330.2+1.0. The component retrieved by pGMCA that we identify as thermal emission is shown on the left. In the middle, we show the component retrieved by pGMCA that we identify as synchrotron emission. On the right, we show an {\it XMM-Newton} three-color image of G330.2+1.0, with 0.4–1.2 keV emission in red, 1.2–2.0 keV emission in green, and 2.0–7.0 keV emission in blue. The contours of our thermal emission component are overlaid in white; they appear in the green clump region that was studied in \cite{Williams_2018} and shown to harbor thermal emission. The scales are logarithmic.}
\label{fig:maps-G330}
\end{figure*}

To study G330.2+1.0, we used the {\it XMM-Newton} 0742050101 observation (see Table \ref{sect:obs}). We reprocessed the data and removed the solar flares using the dedicated {\it SAS} routines, and stacked MOS1, MOS2, and pn observations in a $(x,y,E)$ data cube. We chose a $6$" spatial binning and a $116.8$ eV energy binning to increase the photon counts in each bin. We removed the CCO using an inpainting method.

The pGMCA algorithm was able to retrieve two meaningful components shown in Fig. \ref{fig:maps-G330}. We identify these as synchrotron and thermal emission{, respectively}. The component we identified as thermal was associated with a spectrum endowed with thermal features but too poorly reconstructed to be used as such in a detailed spectral analysis (see the bottom right panel of Fig. \ref{fig:maps-3C58}  for an example of spectra retrieved by pGMCA). Figure \ref{fig:maps-G330} also shows a three-color image of G330.2+1.0 on which the contours of our thermal emission are overlaid. It appears that these contours highlight the region shown in Fig. 4 (east region) of \cite{Park_2009}  and  Fig. 1 of  \cite{Williams_2018}, where both studies detected thermal emission. We did not carry out any spectral analysis of this region as this has already been done extensively  by \cite{Williams_2018}. 

The pGMCA algorithm did not find any other meaningful component that could be interpreted as thermal emission, and we observed that the spatial distribution of the thermal and synchrotron emissions are clearly anti-correlated. Furthermore, the synchrotron component appears slightly recessed around the thermal region, which supports the interpretation of \cite{Williams_2018}  that most of the blast wave is encountering very low-density material, while a small section
encounters a denser region of either the interstellar medium (ISM) or a clump of circumstellar material (CSM). The results obtained using pGMCA on this remnant can be seen as a methodological test as they confirmed past studies, without significantly expanding on them. The major novelty here would be a nondetection: there does not seem to be any other thermal emission than that in the eastern region, which would favor the interpretation that we are observing a clump of ISM or CSM.


\section{3C58}
\label{sect:3C58}

To study 3C58, we used {\it Chandra} 728, 3832, 4382, and 4383 {\it ACIS} observations (see Table \ref{sect:obs}). We reprocessed the data with {\it CIAO v4.15}, and stacked the observations in a $(x,y,E)$ data cube. We chose a $30$" spatial binning and a $43.8$ eV energy binning to increase the photon counts in each bin. We removed the pulsar using an inpainting method. We used {\it Chandra} observations rather than {\it XMM} because the better resolution allowed a spatial rebinning to increase the statistics, without losing too much morphological information. 

\begin{figure*}[ht!]
\centering
\subfloat{\includegraphics[width = 18cm]{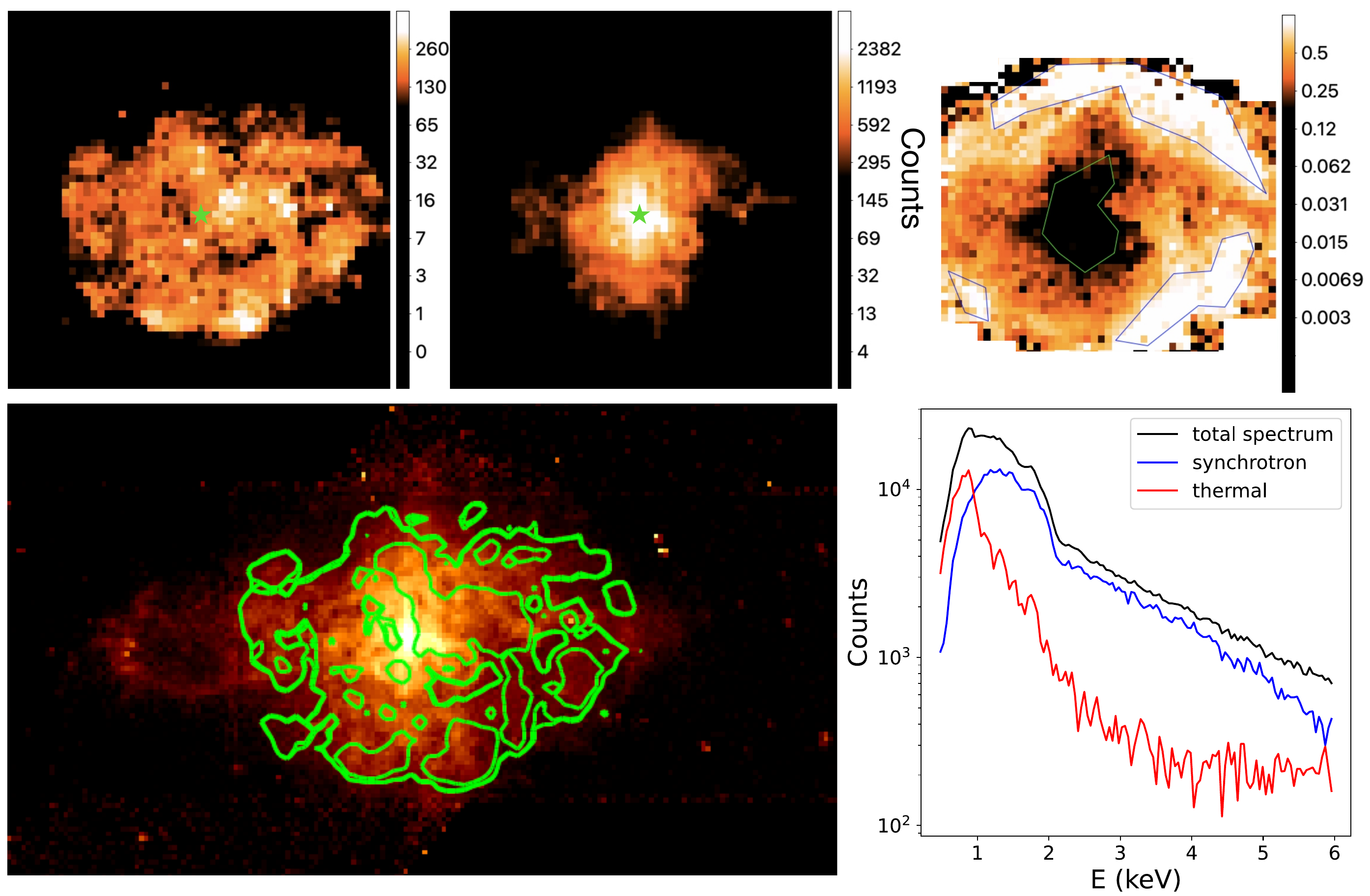}}
\caption{Thermal emission distribution in 3C58. In the top left panel, we show the component retrieved by pGMCA that we identify as thermal emission. In the middle, we show the main component retrieved by pGMCA that we identify as synchrotron emission. The green stars highlight the pulsar position. On the right, we present a ratio map $thermal/(thermal+synchrotron)$. The regions we used to extract a spectrum are overlaid in blue (thermal) and green (synchrotron). In the bottom left, we show a broadband flux image of 3C58 obtained using the \texttt{merge\_obs} routine from CIAO, which is superimposed with the contours from the thermal emission. For all the images, the scales are logarithmic. In the bottom right panel, we show spectra retrieved by pGMCA associated to the thermal and synchrotron emissions.}
\label{fig:maps-3C58}
\end{figure*}

\begin{figure*}[ht!]
\centering
\subfloat{\includegraphics[width = 18cm]{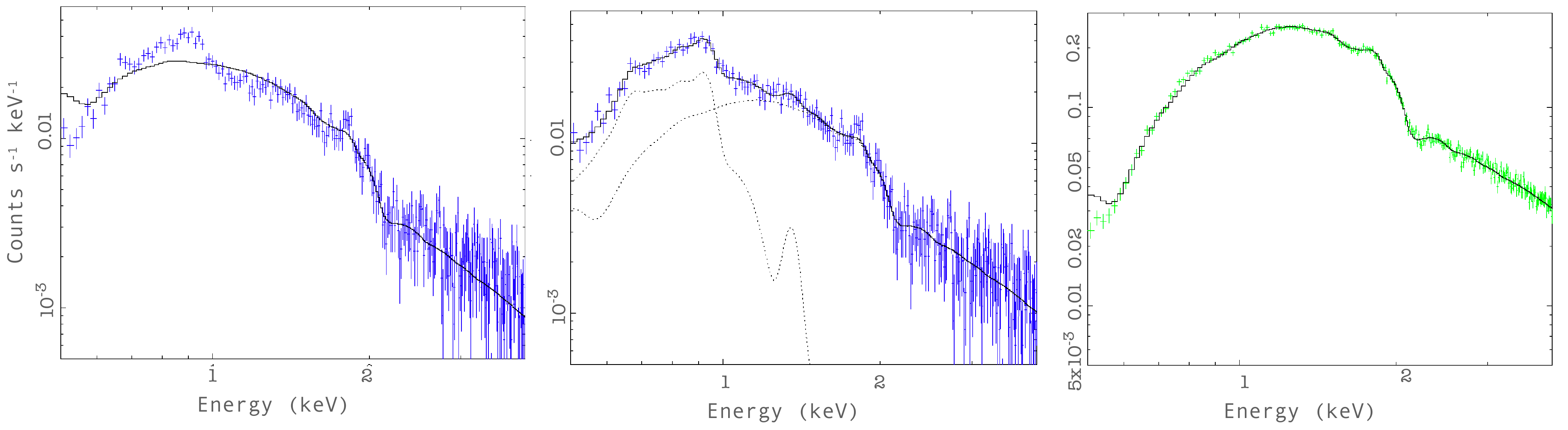}}
\caption{Background-subtracted spectra extracted from the thermal and synchrotron-dominated regions from the right panel of Fig. \ref{fig:maps-3C58}. On the left, we show the spectrum extracted from the thermal region fitted with a \texttt{phabs(powerlaw)} model in {\it Xspec}. In the middle, we show the same thermal spectrum fitted with a \texttt{phabs(powerlaw+vnei)} model. Best-fit parameters of this model are shown in Table \ref{tab:data-description}. On the right, we show the spectrum extracted from the synchrotron-dominated region fitted with a simple \texttt{phabs(powerlaw)} model.}
\label{fig:xspec-3C58}
\end{figure*}

The pGMCA algorithm was able to retrieve two meaningful components shown in Fig. \ref{fig:maps-3C58} that we identify respectively as synchrotron and thermal emission. The component we identified as thermal was associated with a spectrum endowed with thermal features, as can be seen in the bottom right panel of Fig. \ref{fig:maps-3C58} . The global shape is consistent with a thermal spectrum, but it is too poorly reconstructed to be used as such in a detailed spectral analysis. To assess the thermal nature of the latter, we therefore defined extraction regions  by eye based on the ratio map presented in the top right panel of Fig. \ref{fig:maps-3C58}. The resulting background-subtracted spectra are presented in Fig. \ref{fig:xspec-3C58}. 

\begin{table}
    \centering

    \renewcommand{\arraystretch}{1.4}
    \begin{tabular}{c c c}

Model & Parameter & Best fit 
 \\ 
\hline
\texttt{phabs} & nH ($10^{22}$ cm$^{-2}$) & 0.42 $\pm$  0.07 
\\ 
\hline
\texttt{powerlaw} & PhoIndex & 2.55 $\pm$ 0.05 \\ 
  & norm & 8.95E-05 $\pm$ 3.3E-06 \\ 
\hline
\texttt{vnei} & kT (keV) & 0.25 $\pm$ 0.06 \\ 
  & Tau (s/cm$^3$) & 2.4E+11 $\pm$ 4.5E+11 \\ 
  & norm & 1.41E-04 $\pm$ 1.5E-04 \\ 
  & O & 0.583 $\pm$ 0.29 \\ 
  & Ne & 1.84 $\pm$ 0.66\\ 
  & Mg & 1.43 $\pm$ 0.63 \\ 
\hline 
\end{tabular}
\caption{\label{tab:data-description}Best fits obtained in {\it Xspec} to describe the thermal-dominated spectrum from the central panel of Fig. \ref{fig:xspec-3C58}, with a \texttt{phabs(powerlaw+vnei)} model.}
\label{sect:fit}
\end{table}

We used the spectral fitting package {\it Xspec} to fit the spectrum of the  thermal candidate with a \texttt{phabs(powerlaw+vnei)} model, and compared the results with a simple \texttt{phabs(powerlaw)} model, as shown in Fig. \ref{fig:xspec-3C58}. This spectrum clearly contains thermal emission, and the results of our best fit for the \texttt{phabs(powerlaw+vnei)} model are shown in Table \ref{sect:fit}. Similarly, fitting a \texttt{phabs(powerlaw)} model on the synchrotron spectrum confirms the spatial distribution of the thermal and synchrotron components disentangled by pGMCA.

Our results are consistent with the studies from \cite{Gotthelf_2007} and \cite{Slane04}, where an overabundance of Ne IX was detected. Our thermal map also appears similar to the synchrotron-subtracted image shown in Fig. 8 of \cite{Gotthelf_2007}. Using a BSS method allowed the disentanglement of a crisper, cleaner image of this faint thermal component surrounding the core of the synchrotron emission.

The nature of this thermal component has been discussed in previous studies. \cite{Slane04} hypothesized that the overabundance of Ne indicates that this component is composed of ejecta, but \cite{Gotthelf_2007} argue that the solar abundance of Ne has been systematically underestimated by a factor of two \citep{Cunha_2006}. \cite{Bietenholz13} propose two arguments against the thermal emission being ejecta based on results from \cite{Gotthelf_2007} that are consistent with the findings of the present study. The geometrical center of the thermal component from \cite{Gotthelf_2007} is offset from the current pulsar position and its projected origin, and the thermal emission ($\sim 5.6$ pc E-W extent) appears to be smaller than the PWN ($\sim 8.5$ pc E-W extent). While we think that the circular approximation made by \cite{Gotthelf_2007} to describe the  shape of the thermal emission is too coarse and that the argument in favor of a geometrical center is unconvincing as it relies on a spherical evolution assumption, we find the component we extracted to be of a similar $5.6$ pc E--W extent.

The bottom panel of Fig. \ref{fig:maps-3C58} shows the contours  of the thermal emission superimposed on a synchrotron-dominated broadband flux image of the PWN obtained using the \texttt{merge\_obs} routine from CIAO. There is no clear correlation between the thermal component and the filaments from the synchrotron emission, which suggests that there is no interaction between the thermal and nonthermal plasma. The thermal component is therefore likely of ISM origin; it could be ejecta, but only if the reverse shock has not yet reached the center of the remnant.

\begin{figure*}[ht!]
\centering
\subfloat{\includegraphics[width = 18cm]{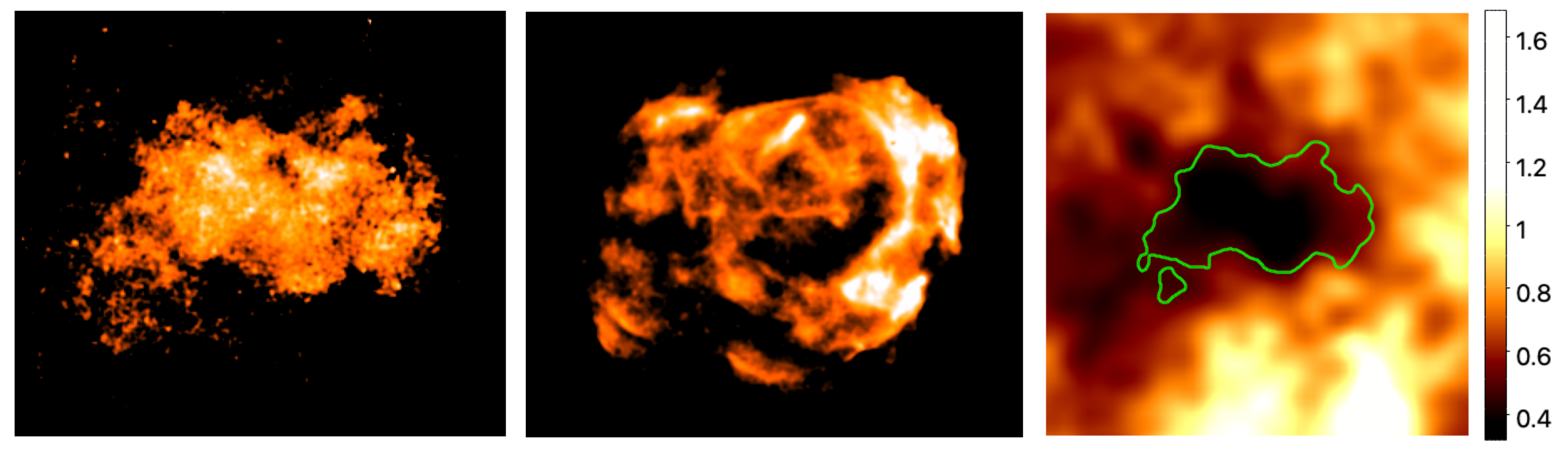}}
\caption{Thermal candidate distribution in \rxj. On the left, we show the component retrieved by GMCA that we identify as potential thermal emission. In the middle, we show the component retrieved by GMCA that we identify as synchrotron emission. The scales are logarithmic. The compact object in the center has been masked. On the right, we show the column density  map derived from optical extinction (in units of $10^{22}$ cm$^{-2}$), over which we superimpose the contour of our thermal candidate.}
\label{fig:maps-RXJ1713}
\end{figure*}

\section{\rxj}
\label{sect:RXJ}

\rxj is a large SNR with $\sim$1
$^{\circ}$ diameter, meaning a collection of {\it XMM-Newton} observations (see Table \ref{sect:obs}) is required to map the entire remnant.
The data were processed differently from the two previous targets presented in this study. These deep observations were combined to produce a mosaic data cube with fine energy binning. The energy range $0.5$ - $10$ keV was divided into 18 energy bands in logarithmic space resulting in energy bins of $\sim$100 eV below 1 keV and $\sim$500 eV at 3 keV. The spatial bin size is of 16". Each energy band gave a distinct image that was processed individually in a similar way to that described in \citet{acero09}, except for the fact that filter wheel closed background files\footnote{\url{https://www.cosmos.esa.int/web/xmm-newton/filter-closed}} were used instead of blank sky event files.
For each observation, for each instrument, and for each energy band, the counts map is subtracted from the filter wheel closed background; the resulting images for all instruments and all observations are then assembled in a mosaic and finally divided by the corresponding mosaic exposure map. The final cube is reassembled using the processed images as "energy slices". Contrary to the data cubes used to study 3C58 and G330.2+1.0, the cube resulting from this process is in flux units instead of counts. For that reason, we used the GMCA algorithm instead of pGMCA, as this latter is only suited for Poissonian data sets. 

Faint thermal emission was previously reported by \citet{Katsuda15} in the center of the SNR where deep exposures dedicated to the compact object were available. If this thermal emission were part of a larger structure, we might expect the GMCA to disentangle the synchrotron-dominated part from the thermal component in a full coverage data set.
The GMCA algorithm was indeed able to retrieve two components at lower energies shown in Fig. \ref{fig:maps-RXJ1713}, which we identify respectively as synchrotron and potential thermal emission. The spectrum of the potential thermal emission  only consists of a small peak at lower energies. In this case, the identification of the fainter component with thermal emission is not obvious, as a spectral analysis similar to the one we conducted for 3C58 did not lead to robust results. In the spectrum extracted from the central region, which is the brightest in our thermal candidate/synchrotron ratio map, we found faint traces of thermal emission at low energies, as reported by \cite{Katsuda15}.

We note that \cite{Pannuti_2003} also claimed to have detected traces of thermal emission in the diffuse emission from the remnant but not in the  bright northwestern rim. 
Our thermal candidate does not show any emission in this region either (see Fig. 1 of \cite{Pannuti_2003}). Similar findings of softer emission in the center than to the northwest were reported by \cite{cassamchenai04}. However, spectra extracted from regions where our thermal candidate is supposed to be present but where the ratio is lower than in the center do not show significant traces of thermal emission. The identification of our thermal candidate is therefore uncertain. 

The fact that the brightest part in our thermal candidate appears at low energies in the only region where thermal emission has been observed with certainty favors its identification as a thermal component. However, our thermal candidate could also be synchrotron emission endowed with an absorption column density and/or spectral index sufficiently different from the main synchrotron component to be separated from it by the algorithm as a spectrally and morphologically distinct source. To test the hypothesis of a spatially different absorption effect, we built a map of the X-ray absorption $N_{\rm H}$ using the optical extinction map $A_{\rm V}$  from \citet{Dobashi_2005}. We used a conversion factor between hydrogen column density and optical extinction $N_{\rm H}$ (cm$^{-2}$) = $2.21 \times 10^{21} A_{\rm V}$ \citep{Guver_2009}.
We are aware that the absorption map derived from extinction has several limitations; one being that the $N_{\rm H}$ map is not restricted to the SNR distance. Nonetheless, \citet{Sano_2015} showed that there is a good correlation between the measured X-ray $N_{\rm H}$ and the optical extinction (see Fig. \ref{fig:maps-RXJ1713}).

The resulting absorption map is shown in the right panel of  Fig. \ref{fig:maps-RXJ1713},  and shows lower absorption in the center region and higher absorption towards the outer parts of the SNR. This map bears morphological similarities to our thermal candidate map.
Therefore, this component could either be of thermal origin, more visible in the center as it is less absorbed at lower energies, or synchrotron emission endowed with a different spectral index and a lower absorption. Both effects could also play a role, and the GMCA might separate a mixed component, containing traces of thermal and nonthermal emission. Further studies would be necessary to assess the  origin of this component with more confidence.

\section{Conclusions}
\label{sect:conclusions}

The use of the GMCA and pGMCA algorithms allowed a detailed mapping of thermal emission in the synchrotron-dominated remnants G330.2+1.0 and 3C58, and led to the discovery of a thermal candidate in \rxj.

In G330.2+1.0, the thermal emission appears localized in the same region studied in \cite{Williams_2018}, supporting the idea that most of the blast wave is encountering very low-density material, while a small section
encounters a denser region of either the ISM or a clump of CSM.

In 3C58, the thermal emission surrounds the X-ray-bright central region, but appears significantly smaller than the PWN ($\sim 5.6$ pc against $\sim 8.5$ pc E--W extent). The nature of this emission has been discussed in previous studies, and these latter mostly consider it to be ISM rather than heated ejecta. Our results are consistent with this conclusion.

In \rxj, the GMCA algorithm retrieved a component in lower energies that could be thermal emission, but we cannot identify this with certainty. The brightest region was shown by \cite{Katsuda15} to harbor spatial emission at lower energies, and its spatial distribution is highly different from that of the synchrotron, favoring a thermal origin.  We find no clear signs of thermal emission in the spectra extracted from other regions.

\begin{acknowledgements}
The material is based upon work supported by NASA under award number 80GSFC21M0002. 
The research leading to these results has received funding from the European Union’s Horizon 2020 Programm under the AHEAD2020 project (grant agreement n. 871158). This work was supported by CNES, focused on methodology for X-ray analysis.
\end{acknowledgements}

%
%
\bibliographystyle{aa} 
\bibliography{thermal-maps}

\appendix

\end{document}